\documentstyle[12pt]{article}
\begin{document}

\title{%
Constraints and Solutions of Quantum Gravity in Metric Representation}
\author{A.\ B\l{}aut\thanks{e-mail address 
ablaut@ift.uni.wroc.pl} and J.\ Kowalski--Glikman\thanks{e-mail 
address 
jurekk@ift.uni.wroc.pl}\\ 
Institute for Theoretical Physics\\ 
University of Wroc\l{}aw\\ 
Pl.\ Maxa Borna 9\\ 
Pl--50-204 Wroc\l{}aw, Poland} 
\maketitle 
 
\begin{abstract} 
We construct the regularised Wheeler--De Witt operator demanding that 
the 
algebra of constraints of quantum gravity is anomaly free. We find 
that 
for a  subset of all wavefunctions being integrals of scalar 
densities this condition can be satisfied. We proceed to finding 
exact solutions of quantum gravity being of the form of functionals of
volume and average curvature of compact three-manifold.
\end{abstract} 
\vspace{12pt} 
PACT number 04.60 Ds 
\clearpage

\section{Introduction} 
 
One of the  outstanding problems of modern theoretical physics is 
the construction of quantum theory of gravity \cite{reviews}. Indeed, 
it have been 
claimed many times that various unsolved problems like the 
cosmological 
constant problem, the problem of origin of the universe, the problem 
of 
black holes radiation will find their ultimate solution once this 
theory 
is finally constructed and properly understood. Some \cite{Emperor}, 
claim that the theory of quantum gravity will also shed some light on 
the 
fundamental problems of quantum mechanics and even on the origin of 
mind. 
These all prospects are very exciting indeed, however, up to now,  
the shapes of the future theory are still very obscured. 
 
Nowadays there are two major ways of approaching the problem of 
quantum gravity. The first one is associated with the broad term 
`superstrings'. In this approach the starting point is a 
two-dimensional quantum field theory which yields quantum gravity as 
a part
of the resulting low-energy  effective theories. It is clear that in 
superstrings, like in other, less 
developed approaches in whose gravity appears as an effective theory, 
it 
does not make sense to try to ``quantize'' classical gravity. 
 
In the 
canonical approach one does something opposite: the idea is to pick 
up 
some structures which appear already at the classical level and then 
promote them to define the quantum theory. In both the standard 
canonical approach in metric representation, which we will follow 
here, 
and 
in the approach based 
on loop variables \cite{loop}, these fundamental structures are 
constraints 
of the  classical canonical formalism reflecting the symmetries and
dynamics of the 
theory,  and their algebra. There are good 
reasons for such an approach. The equivalence principle is the main 
physical principle behind the classical theory of gravity; this 
principle leads to the general co-ordinate invariance and selects the 
Einstein--Hilbert action as the simplest possible one. 
 
Another building block of quantum theory is the quantization 
procedure. Here one encounters the problem as to if a generalisation 
of 
the standard Dirac procedure of quantization of gauge theories in 
hamiltonian
language is 
necessary. This would be the case if one shows that the standard 
approach 
is not capable of producing any interesting results. It is not 
excluded 
that this may be eventually the result of possible failure of 
investigations using standard techniques, however, in our opinion, at 
the moment there 
is no reason to modify the basic principles of quantum theory. 
 
Our starting point consists therefore of 
\begin{itemize} 
\item[(i)] The classical constraints of Einstein's gravity: the 
diffeomorphism constraint generating diffeomorphism of the spatial 
three-surface ``of constant time'' 
\begin{equation} 
{\cal D}_a=\nabla_b\, \pi^{ab} 
\end{equation} 
and the hamiltonian constraint generating ``pushes in time 
direction'': 
\begin{equation} 
{\cal H} = \kappa^2 G_{abcd}\pi^{ab}\pi^{cd} - \frac1{\kappa^2}\sqrt 
h 
(R +2\Lambda) 
\end{equation} 
In the formulas above $\pi^{ab}$ are momenta associated with the 
three-metric $h_{ab}$, 
$$ 
G_{abcd}=\frac1{2\sqrt 
h}\left(h_{ac}h_{bd}+h_{ad}h_{bc}-h_{ab}h_{cd}\right) 
$$ 
is the Wheeler--De Witt metric, $R$ is the three-dimensional 
curvature scalar, $\kappa$ is the gravitational constant, and 
$\Lambda$ 
the 
cosmological constant. The constraints satisfy the following Poisson
bracket algebra 
\begin{equation} 
[{\cal D}, {\cal D}] \sim {\cal D},\label{difdif} 
\end{equation} 
\begin{equation} 
[{\cal D}, {\cal H}] \sim {\cal H},\label{difham} 
\end{equation} 
\begin{equation} 
[{\cal H}, {\cal H}] \sim {\cal D}. \label{hamham} 
\end{equation} 
\item[(ii)] The rules of quantization given by the metric 
representation 
of the canonical commutational relations 
$$ 
\left[\pi^{ab}(x),h_{cd}(y)\right]=-i\delta^{(a}_c\delta^{b)}_d 
\delta(x,y), 
$$ 
$$ 
\pi^{ab}(x)=-i\frac{\delta}{\delta h_{ab}(x)}. 
$$
\item[(iii)] The Dirac procedure according to which one imposes
constraints quantum mechanically by demanding that they (or, better, 
the
corresponding operators) annihilate the subspace of the Hilbert space
called the set of physical states. Bearing in mind the notorious
regularisation and renormalization problems of quantum field theory, 
one
should clearly state what the phrase ``corresponding operators'' 
means.
In general, different choices of such operators could result in
different quantum theories.
\end{itemize}

In the canonical approach, the points (i) to (iii) above 
encompass the whole of the input in our disposal in construction of 
the 
quantum theory. In particular, we do not know what is the correct 
physical inner product, and thus we do not know if the relevant 
operators 
are hermitean or not. Besides, we do not even know if, in the case of
quantum gravity, we should demand 
these operators to be hermitean: the hamiltonian annihilates the 
physical states (the famous time 
problem \cite{ishamtime}) and thus unitary evolution does not play 
any 
privileged role anymore. It follows that, perhaps, we cannot 
distinguish 
``relevant'' 
wave functions by demanding that they are normalizable, as in the 
case 
of quantum mechanics, in fact, since the probabilistic interpretation 
of 
the ``wavefunction of the universe'' is doubtful, it is not clear at 
all if 
the 
norm of this wavefunction is to be $1$. 
 
In the recent paper \cite{JK} a class of exact solutions of the 
Wheeler--De Witt equation was found. In that paper we used the heat 
kernel to regularise the hamiltonian operator and inserted the 
particular operator ordering. The question arises what is the level 
of 
arbitrariness in this construction. In other words, could we 
construct 
other (possibly simpler) regularised hamiltonian operators and what 
would be their 
properties? This question is the subject of the present paper.
\newline
 
It is clear from the discussion above that the only principle, we can 
base our construction on is the principle that the algebra of 
constraints is to be anomaly--free, that is, the corresponding 
algebra 
of commutators of quantum constraints is weakly identical with the 
classical one. This means that the structure of the 
Poisson bracket 
algebra (\ref{difdif}--\ref{hamham}) is to be preserved, in the sense 
which will be explained below, on the quantum level. The following 
section is devoted to the analysis of this problem. 
In section 3 we investigate solutions of the resulting equations. 
Some more technical results are presented in the Appendix.
 
\section{The commutator algebra and construction of regularised 
operators} 
 
As explained in Introduction, our starting point in construction of 
the quantum hamiltonian operator (the Wheeler--De Witt operator) is 
the 
algebra (\ref{difdif}--\ref{hamham}) and we demand that the same 
algebra 
holds on the quantum level. At this point one should ask the question
why we impose such a condition. One of the possible answers is that
the closure
of the algebra is the only principle which makes it possible to find
operators corresponding to classical constraints, required by the 
Dirac
procedure. Another 
argument to be found in the literature is that if the algebra is 
anomalous (that is, if there are additional 
terms resulting from the commutators of the constraint operators) one 
  
cannot find any solutions of
the  quantum constraints. This does not apply 
here since we know explicitly that for a particular
regularisation/regularisation prescription introduced in \cite{JK} a 
class
of solutions exist, and on  solutions the algebra closes identically.
The above argument can be therefore rephrased 
as follows. We want the algebra to close because if it does not, then 
the
solutions we find will have to be in the 
kernel of the anomaly (we know that the kernel is non-empty because 
solutions
do exist) which would mean that 
the conditions we impose will be more restrictive than the ones 
imported from
the classical theory. This by itself is not a disaster, since, in any 
case, 
the classical limit will be the same, but we would like to depart 
from the
classical theory as little as we possibly could. 
 
From our point of view there is another important argument in favour 
of
preserving the constraint algebra 
structure. The vanishing of anomaly is, as it turns out, a quite 
restrictive
condition which makes it possible to 
restrict the form of the employed regulator. The idea is therefore to 
find
a regularisation/renormalization 
procedure consistent with the symmetries dictated by the classical 
general
relativity and to restrict it further by demanding that there are 
solutions
of the theory in a class of natural wavefunctions.

The problem of commutator algebra has been analysed in 
\cite{algebra},  
\cite{algebra1} (and recently in 
\cite{Maeda}) with the result that formal manipulations involving 
point
splitting lead to ambiguous final 
expressions. This conclusion is hardly surprising: It is well known
\cite{Jackiw} that
to compute an anomaly one should first 
define the space of states on which the operator in question act. Then
one should clearly state what is the 
procedure of extracting the finite part of formally divergent 
expressions.
Thus the right question to ask is {\em 
not} what is the formal commutator of constraints but: Given a space 
of
states, does it exist a 
regularisation/renormalization prescription such that the renormalized
action of the operators on the states closes?
It should be stressed that this question is based on the
basic physical interpretation of the relevant operators; indeed,
the constraints operators are  generators of physical transformations
which make sense
only in terms of results of their action 
on appropriate states.  
As it will be seen below,
the condition guaranteeing the absence  of anomalies, not 
surprisingly, 
becomes different when
the operators act on different states. The technical reason is simply 
that
different
states pick up different parts of the regulator. This fact is, of
course, well known in investigation of anomalies in quantum field 
theory
in canonical quantization language (cf.\ \cite{Jackiw}).
\newline

As stressed above, we must start with  choosing the initial 
space of wave functionals. We assume that this space of states is the 
space
of functions of Riemannian functionals, i.e.,
integrals over compact three-space $M$ of scalar densities built of
polynomials in  Ricci tensor, like ${\cal 
V}=\int_M\sqrt h$, ${\cal 
R}=\int_M\sqrt h R$, etc.: 
\begin{equation} 
\Psi = \Psi({\cal V}, {\cal R}, \ldots). 
\end{equation}

We choose the following representation of the diffeomorphism 
constraint 
\begin{equation}
{\cal D}_a(x)=-i\nabla_b^{x} \frac{\delta}{\delta h_{ab}(x)}, 
\label{diffeom}
\end{equation}
where we employed the notation $\nabla_b^{x}$ meaning that the 
covariant 
derivative acts at the point $x$. Then we see that diffeomorphism 
constraint annihilates all the states and the commutator relation 
(\ref{difdif}) is identically satisfied. This is the reason for a
particular, natural ordering in (\ref{diffeom}). Moreover we see that 
the 
relation 
(\ref{difham}) reduces to the {\em formal} relation 
\begin{equation} 
{\cal D}({\cal H}\Psi)\sim {\cal H}\Psi.\label{difham1} 
\end{equation} 
 
Now we must turn to the heart of the problem, the construction of the 
Wheeler--De Witt operator. It is well known that second functional 
derivative acting at the same point on a local functional 
produces divergent result. We 
deal with this problem by making the point split in the kinetic term, 
to wit 
$$ 
G_{abcd}(x)\pi^{ab}(x)\pi^{cd}(x) \Longrightarrow 
\int\, dx'\, K_{abcd}(x,x';t) 
\frac{\delta}{\delta h_{ab}(x)}\frac{\delta}{\delta h_{cd}(x')}, 
$$ 
where $K_{abcd}(x,x';t)$ satisfies 
$$ 
\lim_{t\rightarrow 0^+}K_{abcd}(x,x';t)= G_{abcd}(x')\delta(x,x'). 
$$ 
By virtue of the correspondence principle, we take 
\begin{equation} 
K_{abcd}(x,x';t)=G_{abcd}(x')\triangle(x,x';t)\left(1 + 
K(x,t)\right), 
\end{equation} 
where
$$ 
\triangle(x,x';t)=\frac{\exp\left(-\frac1{4t} 
N_{ab}(x)(x-x')^a(x-x')^b\right)}{4\pi t^{3/2}} 
$$ 
and $K(x,t)$ is a power series in $t$ vanishing at $t=0$.
Using the fact that $t$ has dimension $m^{-2}$ we make the following
expansion for $K$ and $N_{ab}$ 
\begin{equation} 
K(x,t)=a_0tR+(a_1 R^2 + b_1 R_{ab}R^{ab}) + \ldots, \label{kexp}
\end{equation}
\begin{equation}
N_{ab}(x) = h_{ab} + 2t(A_0 R_{ab} + B_0 h_{ab}R) 
+ t^2( B_1 R_a^cR_{cb} + A_1 R_{ab} R + C_1 h_{ab} R^2) + \ldots, 
\label{nexp}
\end{equation} 
where \ldots denote the higher order terms which will not concern us,
and $a$, $b$,  $A$, $B$, $C$ are the free parameters to be fixed.

Next we must resolve the ordering ambiguity in the operator 
${\cal H}$. To 
this end we add the new term $L_{ab}(x)\frac{\delta}{\delta 
h_{ab}(x)}$, 
where  
\begin{equation} 
L_{ab}= \alpha h_{ab} + \beta h_{ab} R + \gamma R_{ab} + \ldots 
\label{lexp}
\end{equation}
contains free coefficients  do be fixed along with the coefficients 
in $K$
and $N_{ab}$. Thus the 
final form of the Wheeler--De Witt operator is\footnote{In the paper 
\cite{JK} we took $\tilde K_{abcd}(x,x')=G_{abcd}(x) \tilde K(x,x')$, 
where $\tilde 
K$ was a heat kernel, and $L_{ab}$ was taken to be the functional 
derivative of $\tilde K_{abcd}$ with respect to $h_{cd}$. } 
$$ 
{\cal H}(x)=\kappa^2\int\, dx'\, K_{abcd}(x,x';t) 
\frac{\delta}{\delta h_{ab}(x)}\frac{\delta}{\delta h_{cd}(x')} + 
$$ 
\begin{equation} 
+ \kappa^2 L_{ab}(x)\frac{\delta}{\delta h_{ab}(x)} 
+ \frac1{\kappa^2}\sqrt h 
(R +2\Lambda).\label{qham} 
\end{equation} 
 
To set the stage, we still need to define the action of operators on 
states. To this end we must discuss the issue of regularisation 
and renormalization. The operator (\ref{qham}) acting on a state 
(defined as an integral of a scalar density) produces, in general, 
terms 
with arbitrary (positive and negative) powers of $t$. This provides 
the 
regularised version of the operator since all the terms are finite, 
and 
singularities of the form $\delta(0)$, $\delta'(0)$, etc.\ are traded 
for
terms which are 
singular for $t\rightarrow0$. Observe that the 
singular 
part of the action of the operator on a state depends on this state.
 To 
renormalize, we follow the procedure 
proposed by Mansfield \cite{Mansfield}, based on analytic 
continuation,
 which result in the following: 
the terms with positive powers of $t$ are dropped, and the singular 
terms of the form $\frac{1}{(4\pi)^{3/2}}t^{-k/2}$ are replaced by the
renormalization 
coefficients $\rho^k$.  Thus we are given a finite 
action of 
the Wheeler--De Witt operator on any state.

There is a number of important comments that must be made at this 
point.
It is
easy to see that the singular part of the regularised action, and thus
the renormalized action of the Wheeler--De Witt
operator on a state does depend on the state (cf.\ (\ref{eqforV}) and
(\ref{eqforR}) in the Appendix.) This is clearly a natural feature of 
any
regularisation technique.   Thus, as already stressed above, the 
condition
for anomaly cancellation should be analysed state-by-state.

Next, the coefficients in the regulator
$K$ are metric dependent. This should not be understood as an 
indication
that the regulator depends on a background metric. It was observed by
many that a wonderful feature of the metric formulation is that the
metric (and its derivatives) appearing in the commutator is to be
understood as result of action
of the metric operator on a state. But in the metric representation 
the
metric acts by multiplication ($\hat h_{ab}(x)| \ast> =  h_{ab}(x)|
\ast>$),  and thus we can always use $h_{ab}(x)$ instead of
$\hat h_{ab}(x)$. Also, the regulator seemingly depends on a 
background
structure through the presence of the explicit $x^a$ terms in the
exponents. Such terms are necessarily present in any regulator based 
on
point splitting technique. However it will be shown below that there 
is
no anomaly in the
quantum mechanical commutator of hamiltonian and diffeomorphism
constraints and this means that the background structure dependence
disappears in the final results.
\newline

Now we can turn to the interpretation of equation (\ref{difham1}).
According to our general philosophy 
explained above, we 
understand it in the following way. A constraint operator acts on a 
state and 
after renormalization gives another state depending on 
renormalization 
constants and the parameters of the regulator. On this resulting 
state the
second operator acts. Thus the 
formal relation (7) is defined to mean (the state $\Psi$ is, by 
definition,
diffeomorphism-invariant)
\begin{equation} 
{\cal D}\,({\cal H}\Psi)_{ren}\sim ({\cal 
H}\Psi)_{ren},\label{difhamf} 
\end{equation} 
and, similarly, for the hamiltonian--hamiltonian commutator 
\begin{equation} 
\left({\cal H}[N]\,\,({\cal H}[M]\Psi)_{ren}\right)_{ren} 
- \left({\cal H}[M]\,\,({\cal H}[N]\Psi)_{ren}\right)_{ren}=0 
\label{hamhamf} 
\end{equation} 
for all $M$ and $N$. In the formula above we used 
the smeared form of the Wheeler--De Witt operator 
$$ 
{\cal H}[M]=\int dx\, M(x){\cal H}(x). 
$$ 
 
Let us turn back to equation (\ref{difhamf}). Since the action of 
diffeomorphism is standard, it suffices to check that $({\cal 
H}\Psi)_{ren}$ is a scalar density. But this is clearly the case: the 
first functional derivative acting on a state produces a tensor 
density 
${\sf T}^{ab}(x')$. After acting by the second derivative and 
contracting indices, we obtain the terms of the form 
$$ 
{\sf T}_0(x')\delta(x',x) + {\sf T}_1(x')\circ \nabla^{x'}\circ 
\nabla^{x'} 
\delta(x',x) + {\sf T}_2(x')\circ \nabla^{x'}\circ \nabla^{x'}\circ 
\nabla^{x'}\circ \nabla^{x'}\delta(x',x) + \ldots 
$$ 
where $\circ$ denotes various indices contractions, and ${\sf T}_n$ 
are 
tensor densities. These terms are multiplied by $\triangle(x,x';t)$ 
and integrated over $x'$. Now we integrate by parts which results in 
replacing 
covariant derivatives acting on $K$ with appropriate powers of 
$t^{-1}$ 
multiplied by some coefficients. After renormalization we obtain a 
scalar density as required. The action of the $L$ term clearly gives 
the 
same result. Thus 
\newline 
 
{\em For the states being integrals of scalar densities there is no 
anomaly in the diffeomorphism --- hamiltonian commutator} 
\newline 
 
This result is quite important because the anomaly in the string 
theory 
appears in the diffeomorphism --- hamiltonian commutator. It proves 
also
that, in spite of implicit co-ordinate system present in the 
construction
of the regulator, the three dimensional diffeomorphisms are not 
broken by
quantum corrections. 
\newline 
 
Now we turn to the most complicated problem, the hamiltonian 
--- hamiltonian commutator (\ref{hamhamf}). Our goal will be to  
use this equation to partially fix the  free coefficients in $K$,
$N_{ab}$, $L_{ab}$. These coefficients will be 
further fixed by demanding existence of solutions of Wheeler--De Witt
equation. In what follows we will be 
interested in solutions of the form $\Psi({\cal V}, {\cal R})$. 
Therefore we check explicitly the closure of the 
algebra only for the states of this form. We will comment on 
the general case at the end of this section.

Let us start with the 
simplest state $\Psi=1$. The action of the first smeared operator 
gives simply 
\begin{equation} 
({\cal H}[M]\Psi)_{ren} = \frac{1}{\kappa^2}\int dx\, \sqrt h(x) M(x) 
(R(x) +2\Lambda). \label{actionofh}
\end{equation} 
Now we have to act on the right hand side of the above equation with 
the operator $({\cal H}[N]$, then 
renormalize the result, and finally subtract the result of the same 
calculation with $N$ interchanged with $M$. 
After rather tedious computation one finds in the commutator the 
term proportional to $[N,M]^a=N\nabla^a M-M\nabla^a M$ which must 
vanish, to wit 
\begin{equation} 
\rho^{(1)}a_0\frac32\nabla_a R - \left(\nabla_a L - 
\nabla_b L_a^b\right) 
=0, 
\end{equation} 
where $L= h_{ab}L^{ab}$. Using Bianchi identity and the expansion 
(\ref{lexp})
we find the first relation 
between coefficients, to wit (it will soon turn out that \ldots terms
in $L_{ab}$ vanish)
\begin{equation}
\frac12 (3\rho^{(1)} a_0 - \gamma) -2\beta=0.\label{coeff1}
\end{equation}
 
Now let us turn to the states depending of ${\cal V}=\int_M d^3x\, 
\sqrt h$.
Let ${\cal H}[M]$ act on this state. From Eq.\ (\ref{eqforV}) we see 
that we
have an equation
which is  of the form
(\ref{actionofh}) (with different coefficients which will include
${\cal V}$.) Therefore
the condition for vanishing of the commutator  is the same as
above, (\ref{coeff1}).

Let us pause for a moment with investigation of the algebra to make an
important observation. We want some $\Psi({\cal V})$ to be a solution 
of
the Wheeler--De Witt equation. From the computations above we see that
the double derivative term in ${\cal H}$ will produce terms up to 
order
$R$. It follows that, while solving the equation, we would not be 
able to
cancel terms of higher
order in $R$ (like $R^2$.) Therefore, all the terms in $L_{ab}$ 
expansion
(\ref{lexp}) denoted by \ldots \  must vanish. Thus we take
 $$ 
L_{ab}=\alpha h_{ab} + (\gamma R_{ab}+\beta h_{ab} R), 
$$ 
where the coefficients $\beta$ and $\gamma$ are subject to the 
condition
(\ref{coeff1}).

Now we turn to the wavefunction $\Psi=\Psi({\cal R})$. 
Let us  analyse the action of the commutator of hamiltonian in a 
number of
steps. The first
observation follows from the $\Psi''$ term in (\ref{eqforR}). It can 
be
checked that after acting on this term by ${\cal H}[N]$ one obtains a
term proportional to $\Psi'''$ which contains unremovable anomaly of 
the
form
$$
[N,M]^a\left(\frac38 R_{ab}\nabla^b R -\frac{3}{16}R\nabla_a R\right).
$$
 Till
this point we assumed that the wavefunction $\Psi({\cal R})$ was
arbitrary, thus anomaly multiplying $\Psi'''$  was to vanish
independently of possible anomalies multiplying different derivatives 
of
$\Psi$. But this is, clearly, cannot be accomplished. We thus have no
choice but
to restrict $\Psi$. It would seem that fixing the background geometry 
\begin{equation}
-\frac38 R^2 + R_{ab}R^{ab}=0
\end{equation}
would do, but this cannot be done because some external condition may 
be
applied only after the commutator is computed, and not at the first 
step.
The only way out is to make $\Psi''=0$, or proportional to $\Psi'$
(where the terms quadratic in curvature are already present.) This 
means
that either $\Psi({\cal R})=A {\cal R}$ or $\Psi({\cal R})$ is a 
linear
combinations of exponents $\exp (\omega {\cal R})$. It should be 
stressed
that since solutions we are after will necessarily have the form of
exponents, the restriction we are making is not as severe as it would
seem at the first sight\footnote{It is sufficient to check that there 
is
no anomaly in an neighbourhood of solutions.}.
The inspection of equation (\ref{eqforR}) clearly shows that the most
economic way is to take ${\cal B}=-\frac38 {\cal J}$. In this way we 
can
cancel the anomaly for arbitrary $\omega$.

It turns out that the anomaly is proportional to
$[N,M]^a$ times a combination of five different tensorial objects. To
cancel the anomaly
proportional to $R_c^b\nabla_b R^c_a$ we must put $B_1=0$. Similarly,
the condition
for the anomaly proportional to $R_c^b\nabla_b R^c_a$ to vanish is 
$b_1=0$.
From the conditions for $R_a^b\nabla_b R$ and $(\nabla_a R)R$, and the
equation ${\cal B}=-\frac38 {\cal J}$  we find
expressions for $A_1$, $C_1$, and $a_1$, to wit
\begin{equation}
A_1= - a_0 A_0 + \frac{11}{4} a_0 + \frac{1}{\rho^{(1)}}\left(
\frac54\gamma + \beta \right),
\end{equation}
\begin{equation}
C_1 = - \frac98 a_0A_0 - \frac{29}{16} a_0 - \frac78
\end{equation}
\begin{equation}
a_1=-\frac{29}{66}a_{0}-\frac{9}{11}a_{0}A_{0}-\frac{1}{\rho^{(1)}}
\left(\frac{14}{11}\beta+\frac{37}{33}\gamma\right)
\end{equation}
There is one equation remaining, being a coefficient of $\nabla_a R$
anomaly which relates ${}_0$ parameters to each other:
\begin{equation}
\rho^{(3)}\left( - \frac{3}{16} - \frac{1}{2} A_0 - \frac{11}{8}B_0 +
\frac{1}{2} a_0 \right) - \frac{3}{2}\alpha =0.\label{0coeff}
\end{equation}
We are left therefore with six free coefficients of the regularised
Wheeler--De Witt operator
$A_0$, $B_0$, $a_0$, $\alpha$, $\beta$, and $\gamma$ subject to two
linear equations (\ref{coeff1}) and (\ref{0coeff}).

Thus the final form of the regularised Wheeler--De  Witt operator 
which
preserves the constraint algebra is (to linear order in $R$ and with
 included)
$$ 
{\cal H}(x)=\kappa^2\int\, dx'\, G_{abcd}(x')\triangle(x,x';t)(1+ a_0 
tR
+ \ldots) 
\frac{\delta}{\delta h_{ab}(x)}\frac{\delta}{\delta h_{cd}(x')} + 
$$ 
\begin{equation} 
+\kappa^2 \left(\alpha h_{ab}+ (\beta h_{ab}R 
+ \gamma R_{ab})\right)(x) 
\frac{\delta}{\delta h_{ab}(x)}+ 
 \frac1{\kappa^2}\sqrt h 
(R +2\Lambda).\label{qhamf} 
\end{equation}

The formula (\ref{qhamf}) completes our construction of the 
Wheeler--De Witt operator. As compared to the choice made in the 
paper 
\cite{JK}, where we used the heat kernel and $L_{ab}$ was its 
functional 
derivative, here we gained much more freedom in the form of additional
free constants. These constant will be further fixed by demanding that
the Wheeler--De Witt equations possesses a maximal number of 
solutions,
that is that there are solutions $\Psi({\cal V})$, $\Psi({\cal R})$, 
and
$\Psi({\cal V},{\cal R})$.

It is possible to extend the above analysis to the states being
functionals of higher powers of curvatures. To this end one has to add
terms of order $t^3$ to the regulator $K$, compute the commutator, and
fix the coefficients as it was done above. It seems quite likely that 
the
resulting equations could be solved. However, the computations are
becoming extremely tedious, and for that reason in this paper we were 
not
able to address the question of anomalies for higher states. 
 
\section{Solutions} 
 
From the previous section we know that the most general form of the 
Wheeler- De Witt operator satisfying our criteria is given by 
equation 
(\ref{qhamf}).
Now, employing this operator, we will try to find a class of 
solutions of the 
Wheeler--De Witt equation. It should be stressed at this point that 
we 
regard the existence of a maximal possible space of solutions as an 
ultimate 
condition fixing the operator completely. The reason for is that any
regulator defines a quantum theory with a certain set of solutions.
Clearly, a theory with  richest set of solutions is most interesting.
 Thus our goal is twofold: to 
find solutions and to fix the operator as to allow for the maximal 
possible number of them.

We will consider  the states of the form $\Psi = \Psi({\cal V}, 
{\cal R})$. It is clear from the form of the Wheeler--De Witt equation
that the resulting equations, multiplying various scalar functions 
will
be linear, therefore, without loss of generality, we can assume that
$ \Psi({\cal V}, 
{\cal R}) = \exp(\sigma {\cal V} + \omega {\cal R})$.

First we solve the equation for the coefficient $\omega$. From the 
part
of the equation involving the square curvature terms,  (\ref{eqforR}),
we see easily that since ${\cal B}= -\frac38{\cal J}$,
$\omega = - {\cal J}=\gamma$. Let
us then turn to the coefficients multiplying $\sqrt h R$ 
\footnote{Observe
that
in addition to terms presented in appendix, (\ref{eqforR}),
(\ref{eqforV}), there is another term resulting from the action of
second functional derivative, one on ${\cal V}$, and one on ${\cal 
R}$. }
\begin{equation}
\kappa^2 \left( \sigma {\cal X} + \omega {\cal Y} -
\frac14\omega\sigma \right) + \frac{1}{\kappa^2}=0,\label{25}
\end{equation}
where
$$
{\cal X} = - \frac{21}{8} a_0 \rho^{(1)} + \frac32\beta + 
\frac12\gamma,\label{26}
$$$$
{\cal Y} = \rho^{(3)}\left( A_0 + 3B_0 - \frac32 a_0+\frac78\right) +
\frac12\alpha. \label{27}
$$
The coefficient multiplying the $\sqrt h$ term reads
\begin{equation}
\kappa^2 \left(- \frac38 \sigma^2  - \frac{21}{8}\sigma\rho^{(3)} +
\frac32 \sigma\alpha - \frac32\rho^{(5)}\omega \right) +
\frac{2\Lambda}{\kappa^2} =0.\label{28}
\end{equation}
\clearpage

{\em Solutions with $\Lambda\neq0$}
\newline

We consider three cases:
\newline

{\em Case I}. $\omega =0$. In this case we have
$$
\Psi=\exp(\tilde\sigma {\cal V}), \;\;\; \tilde\sigma =
-\frac{1}{\kappa^4 {\cal X}},
$$
and equation (\ref{28}) gives a condition for the parameters of the
regulator which can be solved for $\alpha$.

{\it Case II}. $\omega = \gamma$ and $\sigma=0$. We have 
$$
{\cal Y}=-\frac{1}{\kappa^{4}\gamma}
$$
and the condition relating $\gamma$ to the bare coupling and 
renormalization
constants
\begin{equation}
\gamma=\frac{4\Lambda}{3\kappa^{4}}\frac{1}
{\rho^{{(5)}}}.\label{eqforg}
\end{equation}

{\it Case III} $\omega = \gamma$ and $\sigma \neq 0$. We find
$\sigma = 4\alpha-7\rho^{(3)}$ and the condition ${\cal 
X}=\frac{1}{4}\gamma$. This 
condition can be solved along with conditions from the previous 
section
to give expressions for the regulator parameters.

Collecting all results we finally have
\begin{eqnarray}
\Psi_I = \exp\left(-3\frac{\rho^{{(5)}}}{\Lambda}{\cal V}\right) && 
\mbox{Case I};\\
\Psi_{II} = \exp\left(\frac{4\Lambda}{3\kappa^{4}\rho^{{(5)}}}{\cal 
R}\right) && \mbox{Case II};\\
\Psi_{III} = \exp\left(-3\frac{\rho^{{(5)}}}{\Lambda}{\cal V}+
\frac{4\Lambda}{3\kappa^{4}\rho^{{(5)}}}{\cal R}\right) && \mbox{Case 
III}.
\end{eqnarray}
Observe that solution III is a product of solutions I and II. We will
return to this observation below.
\vspace{12pt}

{\em Solutions with $\Lambda=0$}
\newline

it is easy to see that (assuming $\rho^{{(5)}}\neq0$) in this case the
wavefunction $\Psi = \exp(\gamma {\cal R})$ does not solve the
Wheeler--De Witt equation. Taking ${\cal X}=\frac14\gamma$ as above we
find the solution
\begin{equation}
\Psi_{I_0} =\exp\left(-\frac{4}{\kappa^4\gamma}{\cal V}\right)
\end{equation}
The second solution is of the form
\begin{equation}
\Psi_{III_0} =\exp\left(\sigma{\cal V}+ \gamma {\cal R}\right),
\end{equation}
where $\sigma$ is a solution of the following quadratic equation
\begin{equation}
\sigma^2 + \frac{4}{\kappa^4\gamma}\sigma - 4 \rho^{{(5)}}\gamma=0.
\end{equation}
Depending on the value of
$$
\left(\frac{1}{\kappa^4\gamma}\right)^2 + \rho^{{(5)}}\gamma
$$
we have either two real, or two complex, or one real solution.
\newline

Thus we have three different wavefunctions (for both cases $\Lambda 
=0$
and $\Lambda \neq 0$) being 
solutions of the Wheeler--De Witt equation and containing functionals 
of
order at most linear in $R$. It is very interesting that the solutions
depend on the bare coupling constants $\kappa$ and $\Lambda$ and only 
on a
single renormalization constant $\rho^{(5)}$. Of course, any linear
combination (with complex coefficients) of the solution is a solution.
Such combinations will be called below ``Schr\"odinger cat 
universes''.
It can be argued that, contrary to the real solutions, complex
solutions will in general possess a nontrivial time evolution.

\setcounter{equation}{0}
\renewcommand{\theequation}{A.\arabic{equation}}

\appendix
\section{Renormalized action of ${\cal H}[M]$}

Here we present the calculation of the renormalized action of 
hamiltonian
constraint on states. 
We have
$$
G_{abcd}(x')\frac{\delta}{\delta h_{ab}(x)}
\frac{\delta}{\delta h_{cd}(x')}{\cal V}=
-\frac{21}{8}\delta(x-x').
$$
Thus
$$
\int d^3x'\, K(x,x';t)G_{abcd}(x')\frac{\delta}{\delta h_{ab}(x)}
\frac{\delta}{\delta h_{cd}(x')}{\cal V}=$$$$
-\sqrt h\frac{21}{8}\frac{1}{(4\pi)^{3/2}}\left(
\frac{1}{t^{3/2}} + \frac{1}{t^{1/2}}a_0R + O(t) \right).
$$
Using this result and renormalizing, we obtain
$$
{\cal H}[M]\Psi({\cal V})=
\int d^3x\sqrt h M\left\{
-\kappa^2\frac38\Psi''({\cal V})\right. +$$\begin{equation} \left.
+\kappa^2\Psi'({\cal V})\left(
-\frac{21}{8} (\rho^{(3)} + a_0 \rho^{(1)} R) + \frac12 L\right) +
\frac{1}{\kappa^2}(R + 2 \Lambda)
\Psi({\cal V})\right\},\label{eqforV}
\end{equation}
where $L=L_{ab}h^{ab}$.

Similarly,
$$
G_{abcd}(x')\frac{\delta}{\delta h_{ab}(x)}
\frac{\delta}{\delta h_{cd}(x')}{\cal R}=
\frac{7}{8}R(x')\delta(x-x')+ \Box_{x'}\delta(x-x').
$$
Thus
$$
\int d^3x'\, K(x,x';t)G_{abcd}(x')\frac{\delta}{\delta h_{ab}(x)}
\frac{\delta}{\delta h_{cd}(x')}{\cal R}=$$$$
\sqrt h \left[ -\frac32 \rho^{(5)} +\rho^{(3)}R
\left( A_0 + 3B_0 - \frac32a_0+\frac78\right)\right.
+$$$$
+\rho^{(1)}\left\{
        R^2\left( \frac78a_0 +A_1 +3C_1-\frac 32 a_1 +
(A_0+3B_0)a_0 \right)\right. +$$$$+\left.\left.
R_{ab}R^{ab} \left( B_1 - \frac32 b_1 + 3D_1\right)\right\}\right] .
$$
Thus we obtain
$$
{\cal H}[M]\Psi({\cal R})=
\int d^3x\sqrt h M \left\{
\kappa^2\Psi''({\cal R})\left(-\frac38 R^2 + 
R_{ab}R^{ab}\right)\right. +
$$$$ 
+\kappa^2\Psi'({\cal R})\left[
{\cal B} R^2 + {\cal J} R_{ab}R^{ab} +
R\left(\rho^{(3)}\left(A_{0}+3B_{0}-\frac32 
a_{0}+\frac78\right)+\frac12\alpha\right)
-\frac32\rho^{(5)}\right] +$$
\begin{equation}\left.
\frac{1}{\kappa^2}(R + 2 \Lambda)
\Psi({\cal R})\right\},\label{eqforR}
\end{equation}
where
\begin{equation}
{\cal B} = \left[ \frac78a_0 - \frac32 a_1 + (A_0 +3B_0)a_0 
+ A_1 + 3 C_1
\right]\rho^{(1)} + \frac12(\beta + \gamma),\label{calB}
\end{equation}
and
\begin{equation}
{\cal J} =  \left( - \frac32 b_1 + B_1 +3D_1\right)\rho^{(1)} -
\gamma .\label{calJ}
\end{equation}
Equations (\ref{eqforV}) and (\ref{eqforR}) are basic for our
investigations in the main body of the paper. The expressions in the
parentheses $\{\, \ast\, \}$ in these equations are the Wheeler--De 
Witt
equations for the corresponding wavefunctions.

\end{document}